\documentclass[a4paper,11pt]{article}
\usepackage{jinstpub} % for details on the use of the package, please see the JINST-author-manual
\usepackage{lineno}
 \usepackage{float}  
\usepackage{placeins}
\usepackage{subfigure}
\usepackage{caption}     
\usepackage{subcaption}   
\usepackage{textgreek} % allows direct use of Greek letters like γ in text
\usepackage{tabularx}
\usepackage{jinstpub}
\usepackage[font=small,labelfont=bf,justification=raggedright,singlelinecheck=false]{caption}

\title{Performance and long-term aging studies on
Eco-Friendly Resistive Plate Chamber detectors}

\author[1]{D. Ahmadi}
\author[1,2]{M. Tytgat}
\author[4,5]{M. Abbrescia}
\author[21]{G. Aielli}
\author[5,10]{R. Aly}
\author[18]{M. C. Arena}
\author[12]{M. Barroso}
\author[6]{L. Benussi}
\author[6]{S. Bianco}
\author[9]{D. Boscherini}
\author[17]{F. Bordon}
\author[9]{A. Bruni}
\author[19]{S. Buontempo}
\author[17]{M. Busato}
\author[21]{P. Camarri}
\author[14]{R. Cardarelli}
\author[5]{L. Congedo}
\author[12]{D. De Jesus Damiao}
\author[5,8]{M. De Serio}
\author[5]{F. De Bernardis}
\author[21]{A. Di Ciaccio}
\author[21]{L. Di Stante}
\author[15]{P. Dupieux}
\author[20]{J. Eysermans}
\author[3,13]{A. Ferretti}
\author[5,8]{G. Galati}
\author[3,13]{M. Gagliardi}
\author[17]{R. Guida}
\author[4,5]{G. Iaselli}
\author[15]{B. Joly}
\author[24]{S.A. Juks}
\author[4,5]{U. Lakshmaiah}
\author[23]{K.S. Lee}
\author[14]{B. Liberti}
\author[22]{D. Lucero Ramirez}
\author[17]{B. Mandelli}
\author[15]{S.P. Manen}
\author[9]{L. Massa}
\author[5]{A. Pastore}
\author[14]{E. Pastori}
\author[6]{D. Piccolo}
\author[14]{L. Pizzimento}
\author[9]{A. Polini}
\author[14]{G. Proto}
\author[4,5]{G. Pugliese}
\author[3,13]{L. Quaglia}
\author[4,5]{D. Ramos}
\author[17]{G. Rigoletti}
\author[14]{A. Rocchi}
\author[9]{M. Romano}
\author[2]{A. Samalan}
\author[11]{P. Salvini}
\author[21]{R. Santonico}
\author[7]{G. Saviano}
\author[14]{M. Sessa}
\author[5,8]{S. Simone}
\author[3,13]{E. Vercellin}
\author[16]{M. Verzeroli}
\author[22]{N. Zaganidis}

\affiliation[1]{Inter-University Institute for High Energies, Vrije Universiteit Brussel, Pleinlaan 2, 1050 Brussels, Belgium}
\affiliation[2]{Ghent University, Dept. of Physics and Astronomy, Proeftuinstraat 86, 9000 Ghent, Belgium}
\affiliation[3]{INFN Sezione di Torino, Via P. Giuria 1, 10125 Torino, Italy}
\affiliation[4]{Politecnico di Bari, Dipartimento Interateneo di Fisica, via Amendola 173, 70125 Bari, Italy}
\affiliation[5]{INFN Sezione di Bari, Via E. Orabona 4, 70125 Bari, Italy}
\affiliation[6]{INFN - Laboratori Nazionali di Frascati, Via Enrico Fermi 54, 00044 Frascati (Roma), Italy}
\affiliation[7]{Sapienza Università di Roma, Dipartimento di Ingegneria Chimica Materiali Ambiente, Piazzale Aldo Moro 5, 00185 Roma, Italy}
\affiliation[8]{Università degli studi di Bari, Dipartimento Interateneo di Fisica, Via Amendola 173, 70125 Bari, Italy}
\affiliation[9]{INFN Sezione di Bologna, Via C. Berti Pichat 4/2, 40127 Bologna, Italy}
\affiliation[10]{Helwan University, Helwan, Cairo Governorate 4037120, Egypt}
\affiliation[11]{INFN Sezione di Pavia, Via A. Bassi 6, 27100 Pavia, Italy}
\affiliation[12]{Universidade do Estado do Rio de Janeiro, R. São Francisco Xavier 524, 20550-013 Maracanã, Rio de Janeiro - RJ, Brazil}
\affiliation[13]{Università degli studi di Torino, Dipartimento di Fisica, Via P. Giuria 1, 10125 Torino, Italy}
\affiliation[14]{INFN Sezione di Roma Tor Vergata, Via della Ricerca Scientifica 1, 00133 Roma, Italy}
\affiliation[15]{Clermont Université, Université Blaise Pascal, CNRS/IN2P3, Laboratoire de Physique Corpusculaire, BP 10448, F-63000 Clermont-Ferrand, France}
\affiliation[16]{Université Claude Bernard Lyon I, 43 Bd du 11 Novembre 1918, 69100 Villeurbanne, France}
\affiliation[17]{CERN, Espl. des Particules 1, 1211 Meyrin, Switzerland}
\affiliation[18]{Università degli studi di Pavia, Corso Strada Nuova 65, 27100 Pavia, Italy}
\affiliation[19]{INFN Sezione di Napoli, Complesso universitario di Monte S. Angelo ed. 6, Via Cintia, 80126 Napoli, Italy}
\affiliation[20]{Massachusetts Institute of Technology, 77 Massachusetts Ave, Cambridge, MA 02139, USA}
\affiliation[21]{Università degli studi di Roma Tor Vergata, Dipartimento di Fisica, Via della Ricerca Scientifica 1, 00133 Roma, Italy}
\affiliation[22]{Universidad Iberoamericana, Dept. de Fisica y Matematicas, 01210 Mexico City, Mexico}
\affiliation[23]{Korea University, 145 Anam-ro, Seongbuk-gu, Seoul, Korea}
\affiliation[24]{Université Paris-Saclay, 3 rue Joliot Curie, Bâtiment Breguet, 91190 Gif-sur-Yvette, France}

\emailAdd{donya.ahmadi@vub.be}

\abstract{Resistive Plate Chambers detectors are extensively used in several domains of Physics. In High Energy Physics, they are typically operated in avalanche mode with a high-performance gas mixture based on Tetrafluoroethane (C2H2F4), a fluorinated high Global Warming Potential greenhouse gas. The RPC EcoGas@GIF++ Collaboration has pursued an intensive R\&D activity to search for new gas mixtures with low environmental impact, fulfilling the performance expected for the LHC operations as well as for future and different applications. Here, results obtained with new eco-friendly gas mixtures based on Tetrafluoropropene and carbon dioxide, even under high-irradiation conditions, will be presented. Long-term aging tests carried out at the CERN Gamma Irradiation Facility will be discussed along with their possible limits and future perspectives.}

\keywords{Resistive Plate Chambers, Eco-friendly gas mixtures, High Energy Physics detectors, Long-term aging, Gamma Irradiation Facility}

\renewcommand\theequation{\arabic{equation}} 
\usepackage[version=4]{mhchem}

\begin{document}

\maketitle

\flushbottom
\makeatletter
\@addtoreset{equation}{section}
\renewcommand\theequation{\thesection.\arabic{equation}}
\makeatother
\section{Introduction}
\label{sec:intro}
Resistive Plate Chambers (RPCs) are parallel-plate gaseous detectors with high-resistivity electrodes. Given their fast timing, high efficiency, and cost-effective coverage, RPCs are widely used in HEP experiments for muon triggering and tracking  \cite{santonico1981development,cms1997electromagnetic,cortese1999alice,lacava1997atlas}. They are commonly operated in avalanche mode with a mixture of $\mathrm{i\text{-}C_4H_{10}}$ (<5\%), $\mathrm{C_2H_2F_4}$ (>90\%), and $\mathrm{SF_6}$ (<1\%). The latter two components are fluorinated greenhouse gases with high global warming potentials (GWPs) of 1430 and 22800, respectively, and their usage and production are increasingly being restricted in the European Union \cite{EU517,no2024573}. Since 2021, the RPC EcoGas@GIF++ Collaboration \cite{ecogaspaper2024} has been operating RPCs with alternative, more eco-friendly gas mixtures (ECO1\footnote{This mixture was discarded due to relatively high chamber operating voltage and an observed increase in chamber current.}: (45\% HFO / 50\% CO$_2$ / 4\% iC$4$H${10}$ / 1\% SF$_6$), ECO2: (35\% HFO / 60\% CO$_2$ / 4\% iC$4$H${10}$ / 1\% SF$_6$), and ECO3: (25\% HFO / 69\% CO$_2$ / 5\% iC$4$H${10}$ / 1\% SF$_6$)) at the CERN Gamma Irradiation Facility (GIF++), where the chamber performance is evaluated in test beams and possible long-term aging effects are studied under irradiation levels that mimic the HL-LHC conditions \cite{schmidt2016high}. This paper reports on the experimental setup at GIF++, the RPC baseline performance, and the results obtained so far in the long-term aging campaign.

\section{Experimental setup and detectors at GIF++}
\label{sec:exp}
\begin{figure}[H]
\centering
\includegraphics[width=.5\textwidth]{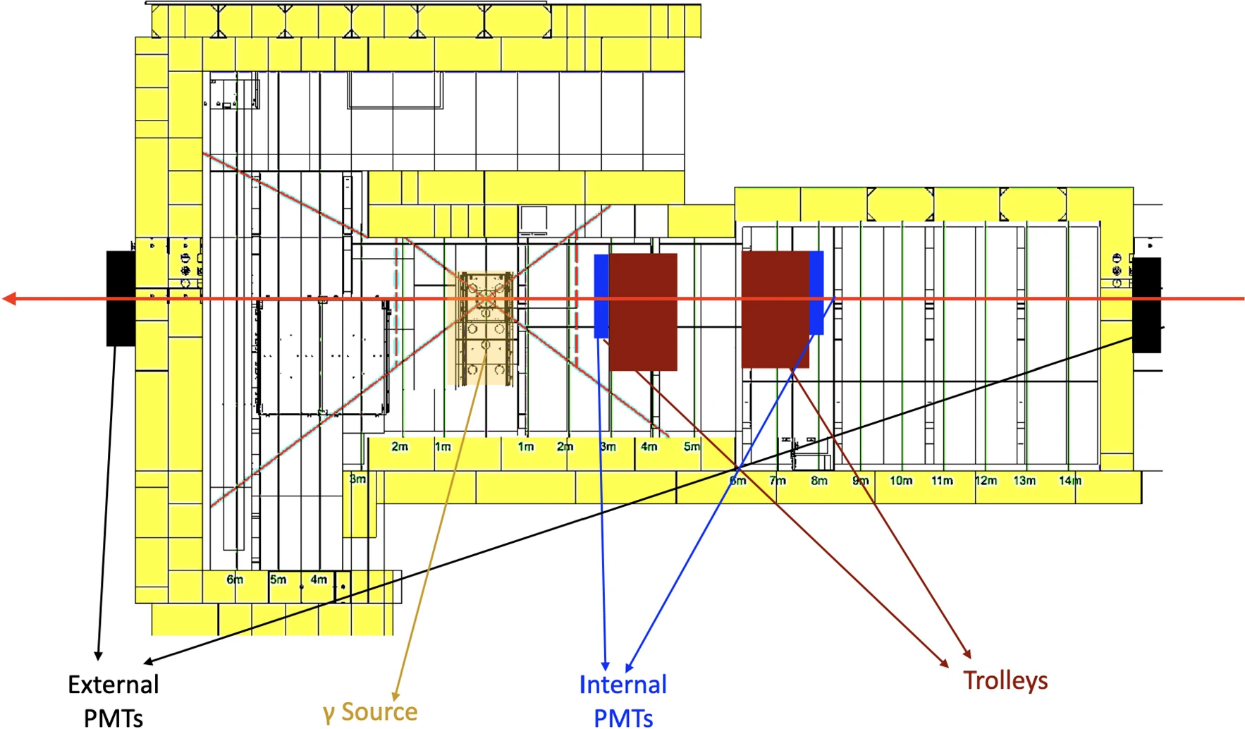}
\caption{Layout of the GIF++ facility showing $^{137}$Cs source (light brown), the two trolleys with test chambers (dark brown), internal scintillators (blue), external scintillators (black), concrete walls (yellow), and the grey outline. The red arrow indicates the muon beam from the H4 line used for the tests.}
\label{fig:experimental_setup}
\end{figure}

% Measurements were performed at the CERN Gamma Irradiation Facility (GIF++), which combines a 12.5 TBq $^{137}$Cs source with a 100 GeV muon beam. This setup enables RPC performance studies under high-rate photon backgrounds, replicating HL-LHC conditions. The intensity of radiation field is regulated by five absorption filters (ABS), while a shaped aluminium filter near the source ensures uniform $\gamma$ flux across vertical planes perpendicular to the beam and bunker walls.
% Detectors from ALICE, ATLAS, CMS, EP-DT, and LHCb/SHiP were installed on movable trolleys at 3 m and 6 m from the source (Table~\ref{tab:chambers}). The LHCb/SHiP chamber was placed at ~6 m in 2021 and ~3 m from 2022 onward. Most chambers were designed as rectangular single-gap structures, while the CMS prototype was a double-gap trapezoid that resembled the  experiment endcaps. Figure~\ref{fig:experimental_setup} shows a simplified GIF++ layout with the trolley positions and main features.
The GIF++ facility combines a 12.5 TBq $^{137}$Cs source with a 100 GeV muon beam. The $\gamma$ radiation field can be tuned via five absorption filters, while an aluminum filter ensures uniform $\gamma$ flux. Detectors from ALICE, ATLAS, CMS, EP-DT, and LHCb/SHiP were placed on movable trolleys at 3 m and 6 m from the source (Table~\ref{tab:chambers}). Figure~\ref{fig:experimental_setup} shows the simplified layout. For muon-beam tests, a $10 \times 10$ cm$^2$ trigger was implemented using four scintillators, two positioned inside and two outside the irradiation zone.

\begin{table}[htbp]
\centering
\caption{Characteristics of the chambers under test.}
\label{tab:chambers}
\resizebox{\textwidth}{!}{%
\begin{tabular}{lcccccc}
\hline
Group & Distance from $^{137}$Cs [m] & Dimensions [cm$\times$cm] & Gap [mm] & Electrode [mm] & \# of strips & Strip pitch [mm] \\
\hline
ALICE & 6.0 & 50$\times$50 & 2.0 & 2.0 & 16+16 & 30 \\
ATLAS & 3.2 & 55$\times$10 & 2.0 & 1.8 & 1 & N/A \\
CMS & 3.5 & (41.5+23.9)$\times$100.5 & 2+2 & 2.0 & 128 & 5.5/11.5 \\
EP-DT & 3.0 & 70$\times$100 & 2.0 & 2.0 & 7 & 21 \\
LHCb/SHiP & 6.2 & 70$\times$100 & 1.6 & 1.6 & 32+32 & 10.6 \\
\hline
\end{tabular}%Layout of the GIF++ facility showing $^{137}$Cs source...
}
\end{table}

%Figure 1

During measurements, the high voltage applied to the RPCs was adjusted for temperature and pressure variations to maintain a constant effective voltage (HVeff) \cite{abbrescia2013operation} using 
%Equation~\eqref{eq:hv_corr}:

% $\mathrm{HV}_{\mathrm{eff}}$

\begin{equation}
\mathrm{HV}_{\mathrm{app}} = \mathrm{HV}_{\mathrm{eff}} \left[ (1 - \alpha) + \alpha \frac{P}{P_0} \frac{T_0}{T} \right],
\label{eq:hv_corr}
\end{equation}

where $P$ and $T$ denote atmospheric pressure and temperature, with reference values $P_0$ = 990 mbar and $T_0$ = 293.15 K, and $\alpha$ is an empirical constant equal to 0.8. Unlike the one used in 2021\cite{abbrescia2024high}, this correction ensures a better long-term stability\cite{abbrescia2013operation}.

As detailed in Table~\ref{tab:electronics}, the RPC readout depended on the experiment that provided the chambers, i.e., ALICE and LHCb/SHiP chambers used FEERIC ASICs with optimized thresholds, CMS prototypes were equipped with their standard front-end, and ATLAS and EP-DT relied on waveform digitizers. The data acquisition was performed with TDCs or digitizers, followed by collaboration-specific analysis frameworks.
\begin{table}[htbp]
\centering
\caption{Electronics used for signal readout}
\label{tab:electronics}

\resizebox{\textwidth}{!}{%
\begin{tabular}{lccc}
\hline
Group & FE chip & FE equivalent threshold & DAQ type \\
\hline
ALICE     & FEERIC ASICs      & 200 mV horiz., 106 mV vert. (after amp.) & TDC CAEN V1190A \\
ATLAS     & N/A               & 130 fC                                  & Digitizer CAEN DT5730 \\
CMS       & CMS/RPC standard  & 220 mV = 150 fC (after amp.)            & TDC CAEN V1190A \\
EP-DT     & N/A               & 2 mV                                    & Digitizer CAEN V1730 \\
LHCb/SHiP & FEERIC ASICs      & 70 mV horiz., 80 mV vert. (after amp.)  & TDC CAEN V1190A \\
\hline
\end{tabular}%
}
\end{table}

\section{RPC performance before aging studies}
\label{sec:perb}

\begin{figure}[htbp]
  \centering
  \subfigure{\includegraphics[width=0.35\linewidth]{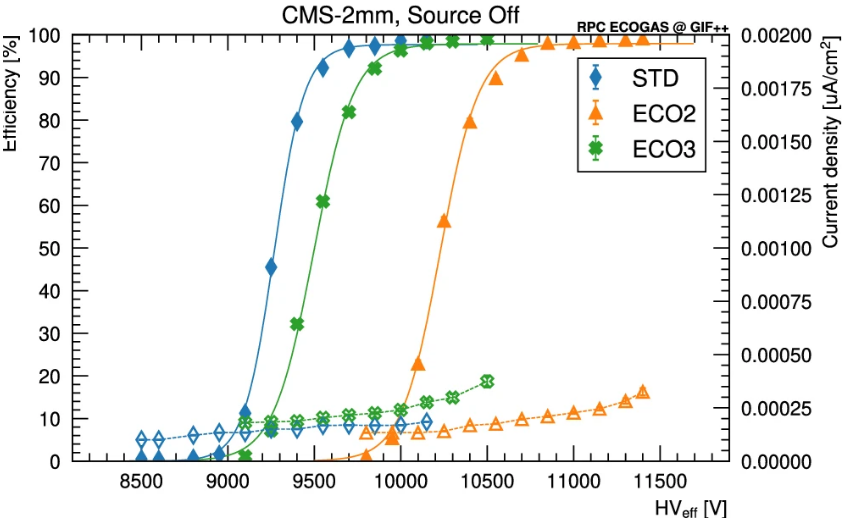}}\label{fig:RE11_b_performance}
  \hspace{0.05\linewidth}
 \subfigure{\includegraphics[width=0.35\linewidth]{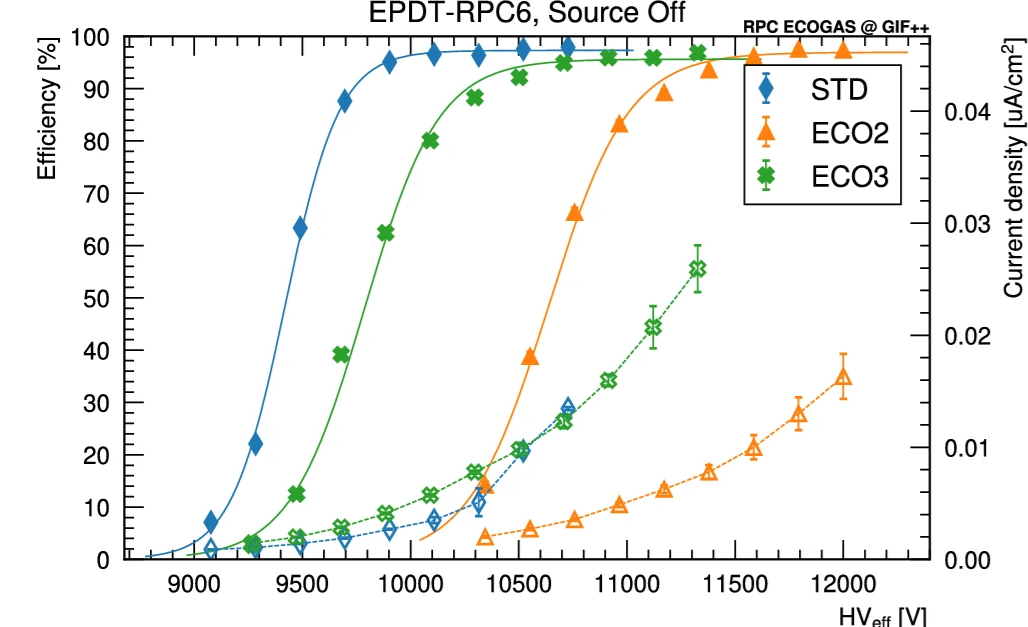}}\label{fig:EPDT_b_performance}
  \caption{Left: Efficiency and current density, as a function of $\mathrm{HV}_{\mathrm{eff}}$, were measured for the CMS chamber without irradiation, sequentially filled with the STD, ECO2, and ECO3 gas mixtures. Right: same, but for the EP-DT chamber.}
  \label{fig:performance_CMS_EPDT}
\end{figure}

The chamber baseline performances without \ce{^{137}Cs} irradiation were evaluated using the STD, ECO2, and ECO3 gas mixtures \cite{abbrescia2024high}. Figure~\ref{fig:performance_CMS_EPDT} shows the RPC efficiency and current density as a function of $\mathrm{HV}_{\mathrm{eff}}$ for the CERN EP-DT and CMS chambers. Both chambers reach efficiency plateaus above 95\%. The CMS chamber exhibits higher efficiency at lower voltages and a steeper rise due to its double-gap design that combines signals on the same readout strips. The CERN EP-DT chamber shows a higher current density, likely due to construction details and prior use in other tests.
%________________________________________________
% figure 2
% %________________________________________

\begin{figure}[htbp]
  \centering
  \subfigure{\includegraphics[width=0.329\linewidth]{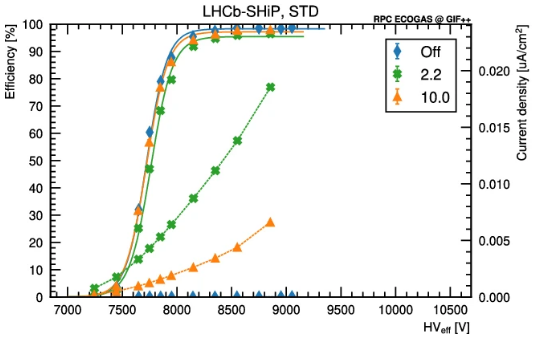}}
  \label{fig:LHCb_std}
  \hfill
 \subfigure{\includegraphics[width=0.329\linewidth]{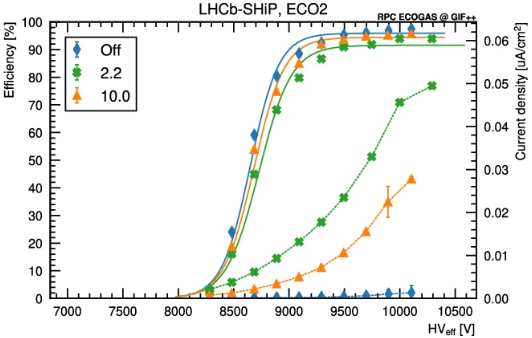}}
 \label{fig:LHCb_ECO2}
  \hfill
 \subfigure{\includegraphics[width=0.329\linewidth]{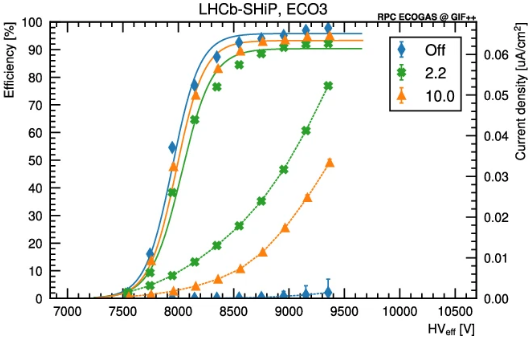}}\label{fig:LHCb_ECO3}
  \caption{Left: Efficiency and current density versus $HV_{\text{eff}}$ for the LHCb/SHiP chamber operated with the STD mixture, obtained without irradiation, and with ABS = 10 and 2.2. Center: Corresponding results for the same chamber filled with the ECO2 mixture. Right: Same, but with the ECO3 mixture.}
  \label{fig:performance_withirra_LHCb}
\end{figure}

Figure~\ref{fig:performance_withirra_LHCb} shows the efficiency and current density as a function of $\mathrm{HV}_{\mathrm{eff}}$ for the LHCb-SHiP chamber under irradiation. The left, central, and right panels correspond to the chamber filled with STD, ECO2, and ECO3 gas mixtures, respectively. Measurements were taken with absorption filters ABS = 10 and 2.2, corresponding to absorbed doses of 510 and 2070 µGy/h. The results obtained without irradiation are superimposed on the plots for reference.
%_______________
% figure 3 

\begin{figure}[htbp]
  \centering
  \subfigure{\includegraphics[width=0.329\linewidth]{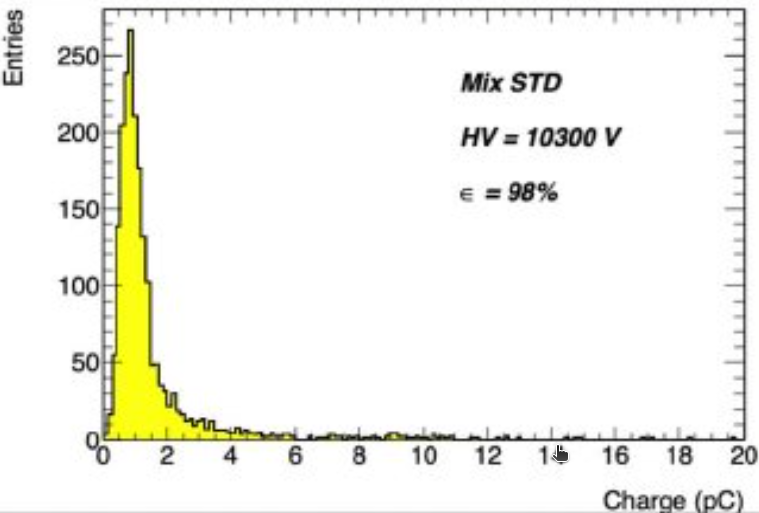}}
  \label{fig:ATLAS_std}
  \hfill
 \subfigure{\includegraphics[width=0.329\linewidth]{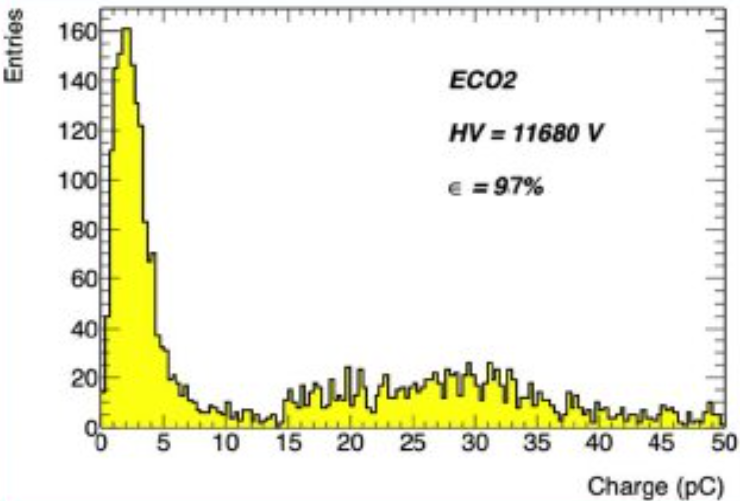}}
 \label{fig:ATLAS_ECO2}
  \hfill
 \subfigure{\includegraphics[width=0.329\linewidth]{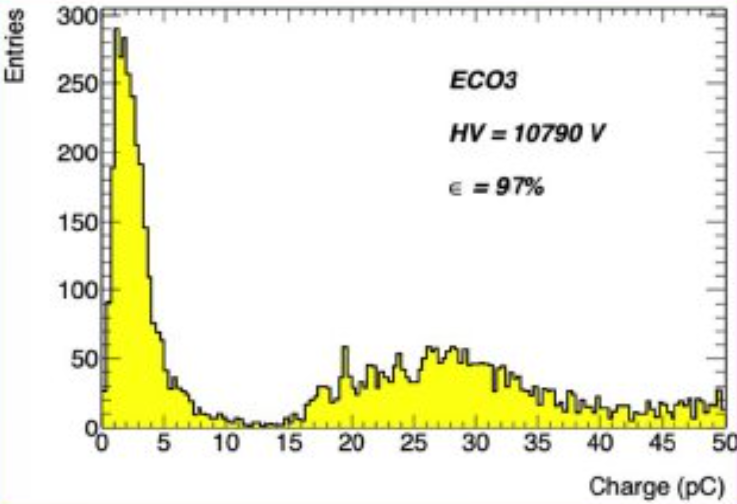}}\label{fig:ATLAS_ECO3}
  \caption{Charge distributions of the signals induced on the 3 cm wide strip of the ATLAS RPC, measured at HVknee , for the
STD, ECO2 and ECO3 gas mixtures, from left to right, respectively.}
  
  \label{fig:side_by_side 6}
\end{figure}

The measurements indicate that as the dose increases, the plateau efficiency decreases, and this effect is stronger for ECO2 and ECO3 mixtures than for STD. Current densities $J_{\text{knee}}$, measured at $HV_{\text{knee}}$ (95\% plateau efficiency), are systematically higher with ECO2 and ECO3 gas mixtures. For single-gap chambers like LHCb-SHiP, the current density increases by about a factor of two with these mixtures. This effect can be attributed to the broader charge distributions that are observed for the ATLAS RPC as shown in Figure~\ref{fig:side_by_side 6}, and is linked to the higher \ce{CO2} content in the eco-friendly gas mixtures. Larger chamber currents will accelerate the RPC aging and should thus be avoided.

% figure 4 

\FloatBarrier
\section{Aging campaign}
\label{sec:agi}

%\subsection{Methodology}
% I have an opinion to put the name of it aging methodology and add another section under the name of aging studies and there I can discuss about the resukts of the current and integrated charge density resulting from the methodology

%\label{sec:Methodology}

Throughout the still ongoing aging test, the chambers are continuously flushed with the gas mixture and operated at a fixed high voltage. The absorbed currents are monitored by the power supplies with a precision of $0.1\mu\text{A}$. To maintain a constant effective voltage, the webDCS data-acquisition automatically applies the temperature and pressure corrections using Equation~\ref{eq:hv_corr}.
%________________________________________________
%figure 5
\begin{figure}[htbp]
  \centering
  \subfigure{\includegraphics[width=0.4\linewidth]{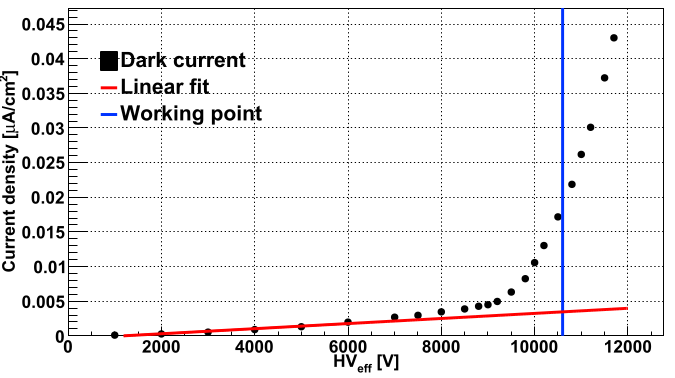}\label{fig:ohmic_current}}
  \hspace{0.05\linewidth}
  \subfigure{\includegraphics[width=0.4\linewidth]{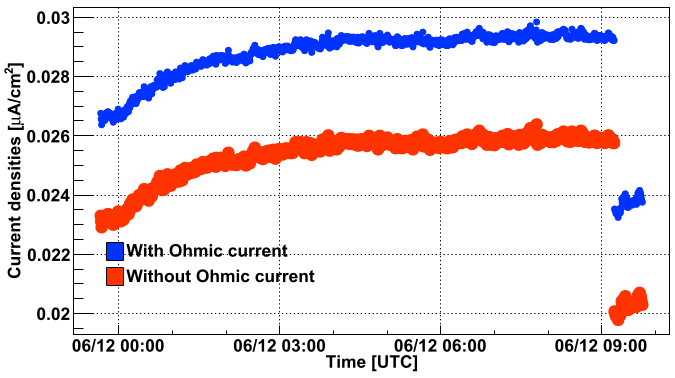}\label{fig:current_with_without}}
  \caption{The left panel shows the dark current density versus $\mathrm{HV}_{\mathrm{eff}}$ for the EP-DT detector. The ohmic component of the dark current is visible, along with the linear interpolation used to derive the ohmic dark current at the irradiation voltage. The right panel presents the current density under irradiation, before and after ohmic subtraction, as a function of time. The step observed near the end reflects a change in irradiation conditions, leading to a current decrease.}
  \label{fig:side_by_side1}
\end{figure}
%________________________________________

Values for the current, effective, and applied voltages are stored every 30 seconds. Once a week, the \ce{^{137}Cs} source is fully shielded to measure the absorbed current without $\gamma$ irradiation. This “dark current” is a key parameter, as any increase may be a sign of detector aging. Figure~\ref{fig:side_by_side1} shows the dark current density versus $\mathrm{HV}_{\mathrm{eff}}$ for the EP-DT detector with the ECO2 mixture. Even at voltages well below the avalanche threshold (7–8 kV for a 2 mm gap), a nonzero current exists, that is identified as the ohmic component, which flows through conductive paths rather than the gas. Weekly source-off measurements allow monitoring this contribution.\\
\indent During irradiation, the chambers are operated at fixed $\mathrm{HV}_{\mathrm{eff}}$ (10.6 kV for 2 mm gaps, 8–9 kV for 1.6 mm with ECO2). The ohmic component is subtracted from the measured total current density, yielding the current density that is flowing through the gas and is associated with the $\gamma$ irradiation.
Figure~\ref{fig:side_by_side1} illustrates the current density absorbed during irradiation for the EP-DT detector. The blue markers show total current density, and red markers the corrected values (the current after the ohmic component was subtracted), which are then used to compute the integrated charge density.

%________________________________________________

\begin{figure}[htbp]
  \centering
  \subfigure{\includegraphics[width=0.4\linewidth]{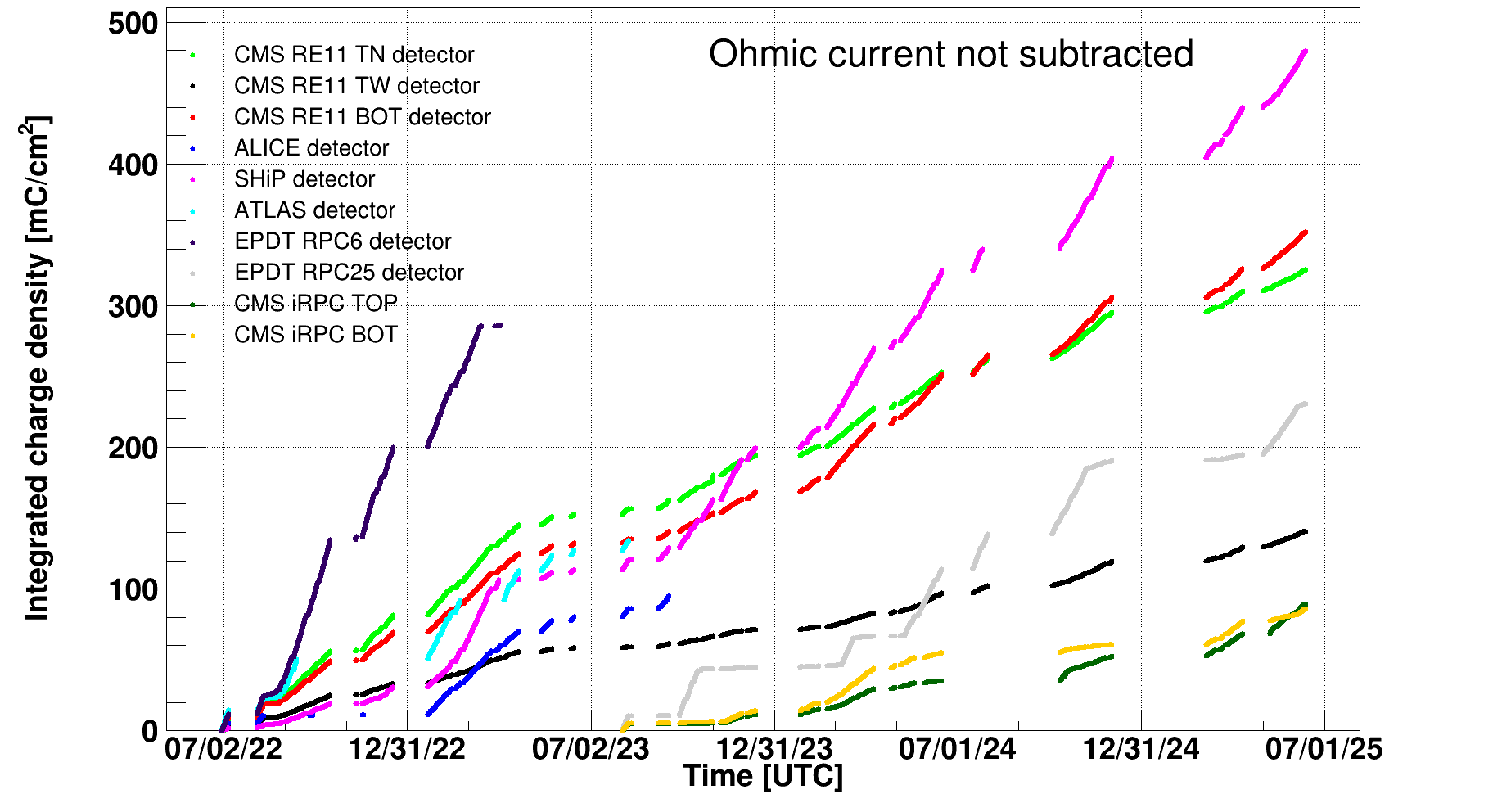}\label{fig:inte_time_No_ohmic_subtracted}}
  \hspace{0.05\linewidth}
  \subfigure{\includegraphics[width=0.4\linewidth]{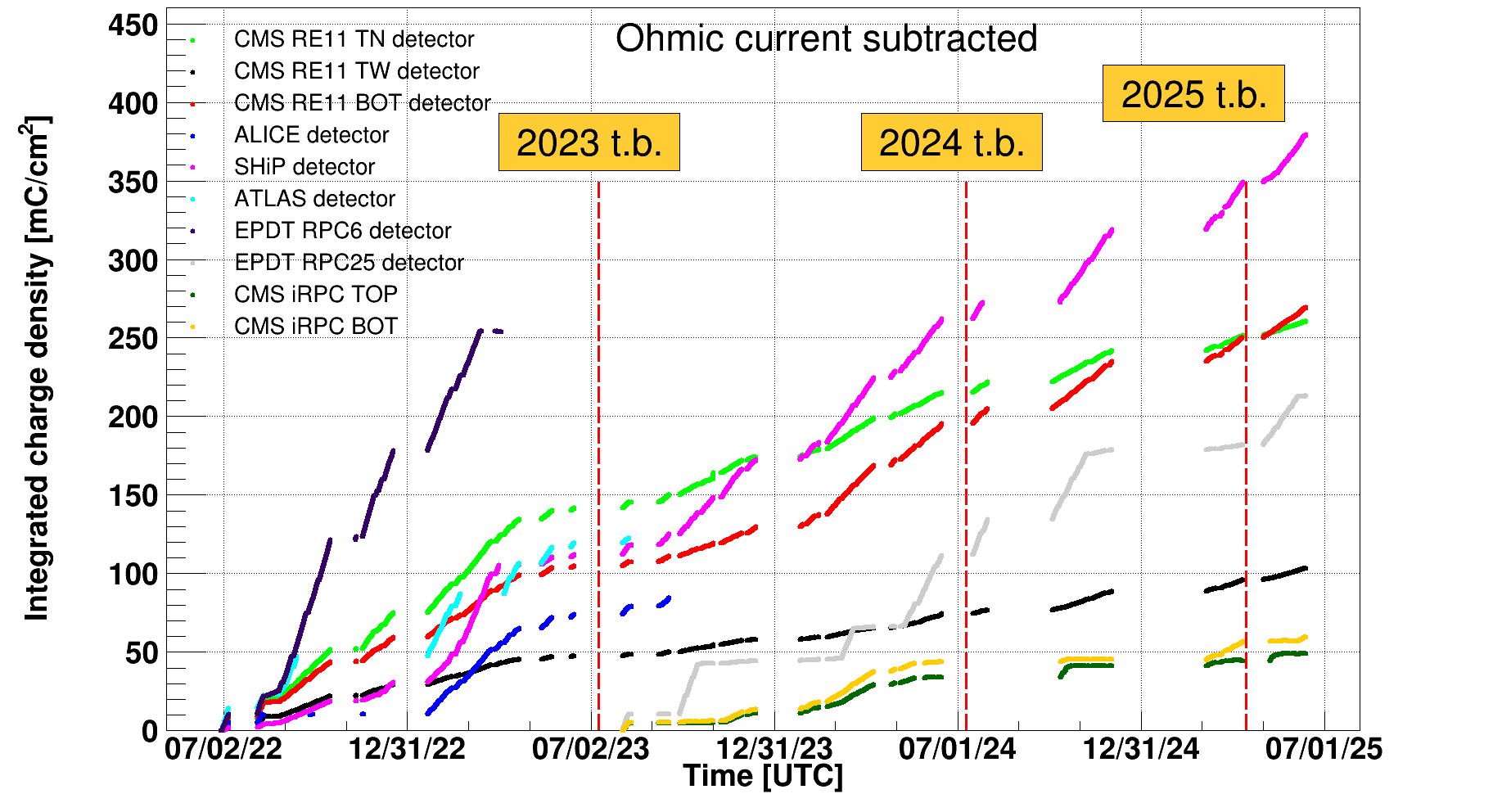}\label{fig:inte_time_ohmic_subtracted}}
  \caption{Time-dependent accumulated charge density for all RPC ECOgas@GIF++ collaboration detectors, i.e., the left panel shows the total integrated current density, while the right panel displays the results after subtracting the ohmic component.}
  \label{fig:side_by_side2}
\end{figure}

%\subsection{Aging studies}
Aging is studied as a function of the integrated charge density, which is defined as the time integral of the current density passing through the detector (mC/\ce{cm^{2}}). Figure~\ref{fig:side_by_side2} shows the accumulated charge density corresponding to irradiation from July 2022 to July 2025, where the gaps in the data correspond to periods without irradiation. The left panel shows the total integrated charges, while the right panel displays the results after subtracting the ohmic component which thus correspond to the charge transported through the gas gap. Differences among detectors can be attributed to the fact that the selected irradiation voltage does not precisely correspond to the same efficiency level. The integrated charge targets also vary per experiment, i.e., ALICE aims for $\sim$100 mC/\ce{cm^{2}}, while CMS requires $\sim$1 C/\ce{cm^{2}} (including a safety factor of three). The detector distance from the source further affects the accumulated charge density, as evident for the ALICE chamber relative to the others. Between July 2022 and July 2025, the detectors were operated with the ECO2 gas mixture, using a source attenuation filter of 2.2 for most irradiation studies.

% Figure 6
%_________________________________________
% The $\mathrm{HV}_{\mathrm{eff}}$ for irradiation was 10.6 kV for 2 mm gap detectors and 8.8–9 kV for the 1.6 mm SHiP RPC.

\begin{figure}[htbp]
  \centering
  \subfigure{\includegraphics[width=0.35\linewidth]{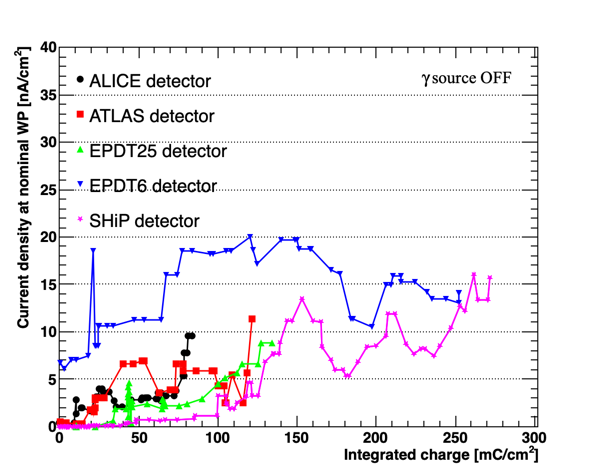}\label{fig:total_alice}}
  \hspace{0.05\linewidth}
  \subfigure{\includegraphics[width=0.35\linewidth]{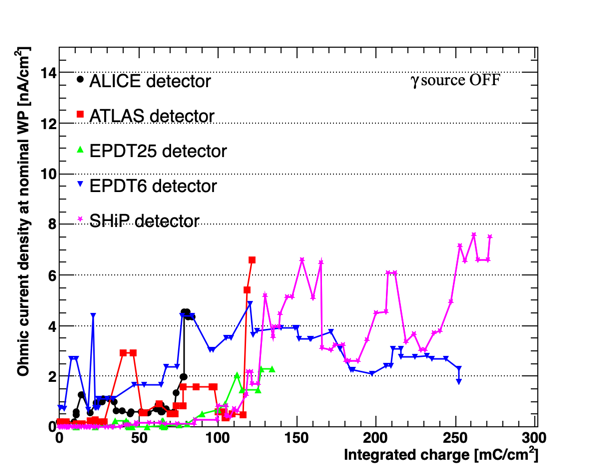}\label{fig:ohmic_alice}}

   % Reduce the vertical gap here
  \vspace{-1em}   % adjust value: -1em, -0.5em, etc.
  
  \subfigure{\includegraphics[width=0.35\linewidth]{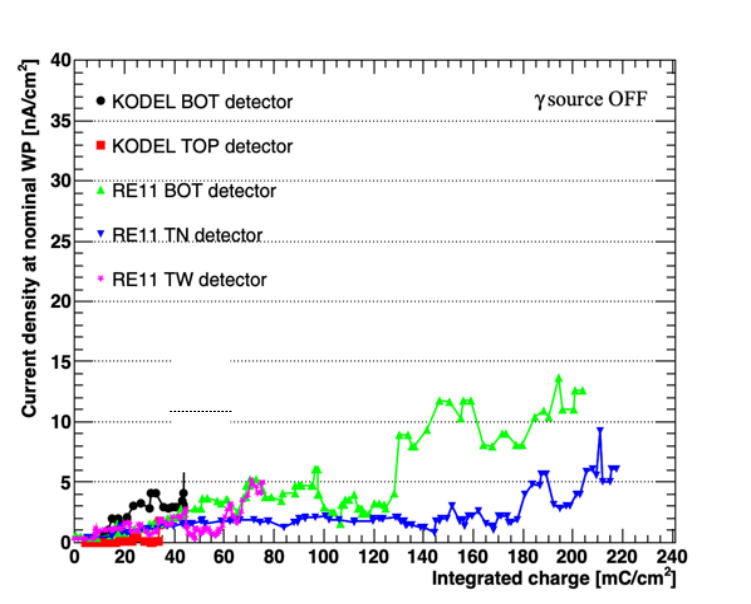}\label{fig:total_RE11}}
  \hspace{0.05\linewidth}
  \subfigure{\includegraphics[width=0.35\linewidth]{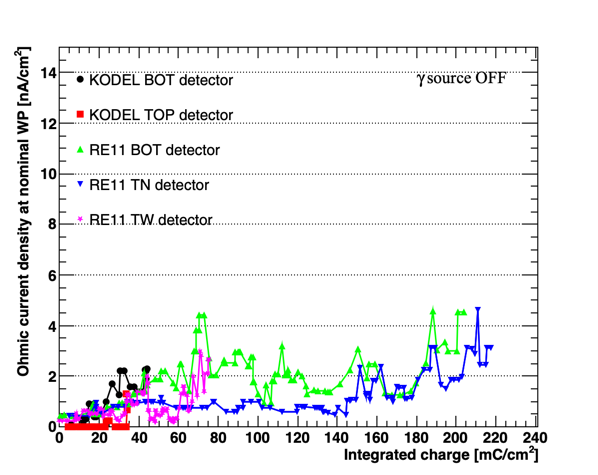}\label{fig:ohmic_RE11}}
  
  \caption{RPC current density at nominal working point as a function of integrated charge, measured with the $\gamma$ source off. The top-left and top-right panels show the total and ohmic current components, respectively, for the ALICE, ATLAS, EPDT, and SHiP detectors. The bottom-left and bottom-right panels present the corresponding results for the KODEL TOP and BOT detectors, as well as the RE11 BOT, TW, and TN detectors.}
  \label{fig:currrent_integrated_four_plots}
\end{figure}

The trend of the absorbed current density throughout the entire aging campaign is illustrated in Figure~\ref{fig:currrent_integrated_four_plots}, which shows the absorbed current density as a function of integrated charge density. The top panels display the total and ohmic current densities for the ALICE, ATLAS, EPDT, and SHiP detectors, while the bottom panels provide results for the KODEL TOP/BOT and CMS RE11 BOT, TW, and TN detectors. In general, an increase in the current is observed after an integrated charge of about 50–100 mC/cm², depending on the detector. The CERN EP-DT detector 6, for instance, had to be replaced by detector 25 in 2022 due to persistently high currents that were already present from the beginning of the aging campaign. Since then, most detectors exhibited current fluctuations followed by a slow but steady rise over time.
% Figure 7
\section{RPC performance after aging}  
\label{sec:pera}
\begin{figure}[htbp]
  \centering
  \subfigure{\includegraphics[width=0.329\linewidth]{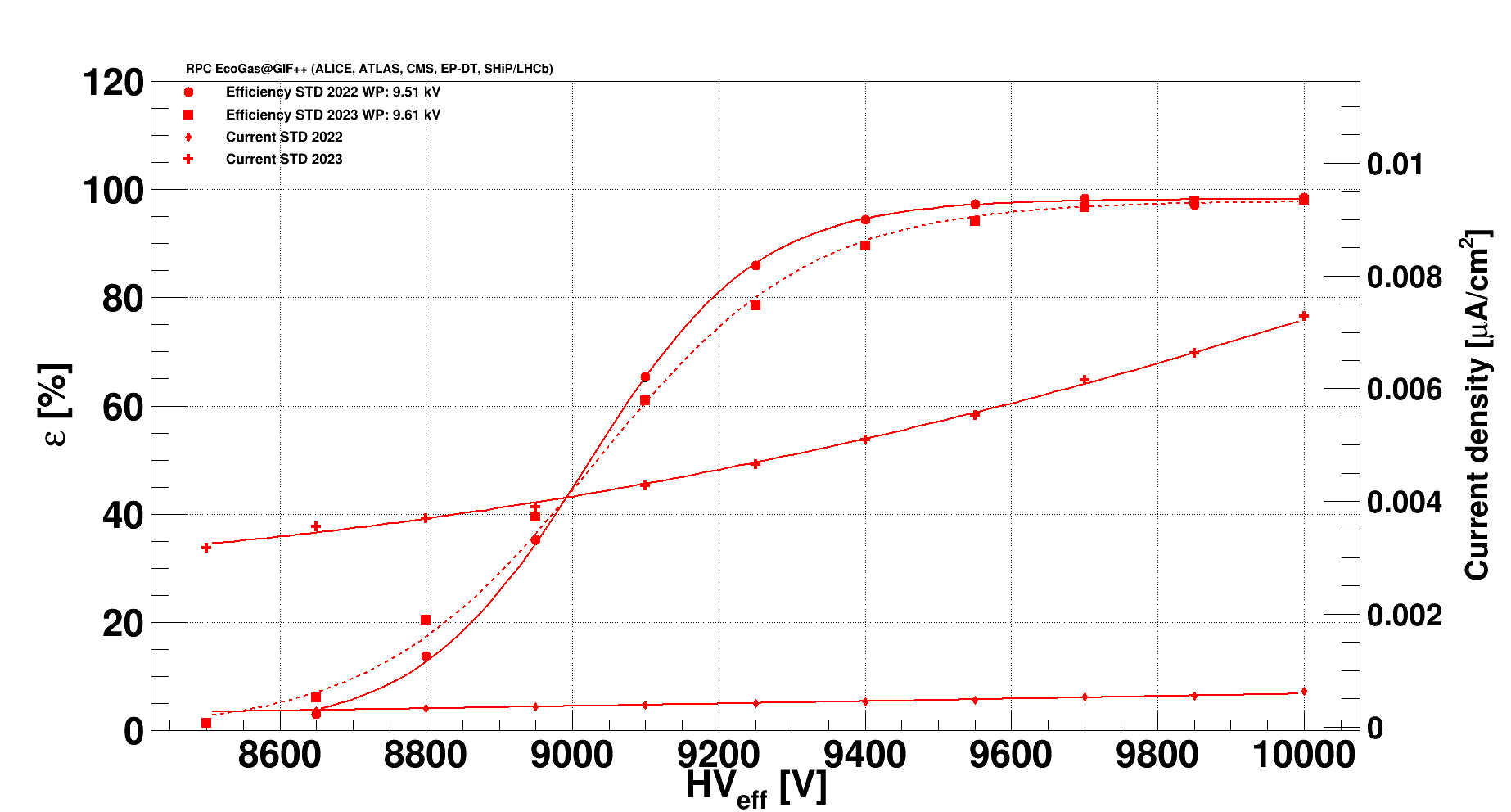}}
  \label{fig:Alice_std}
  \hfill
 \subfigure{\includegraphics[width=0.329\linewidth]{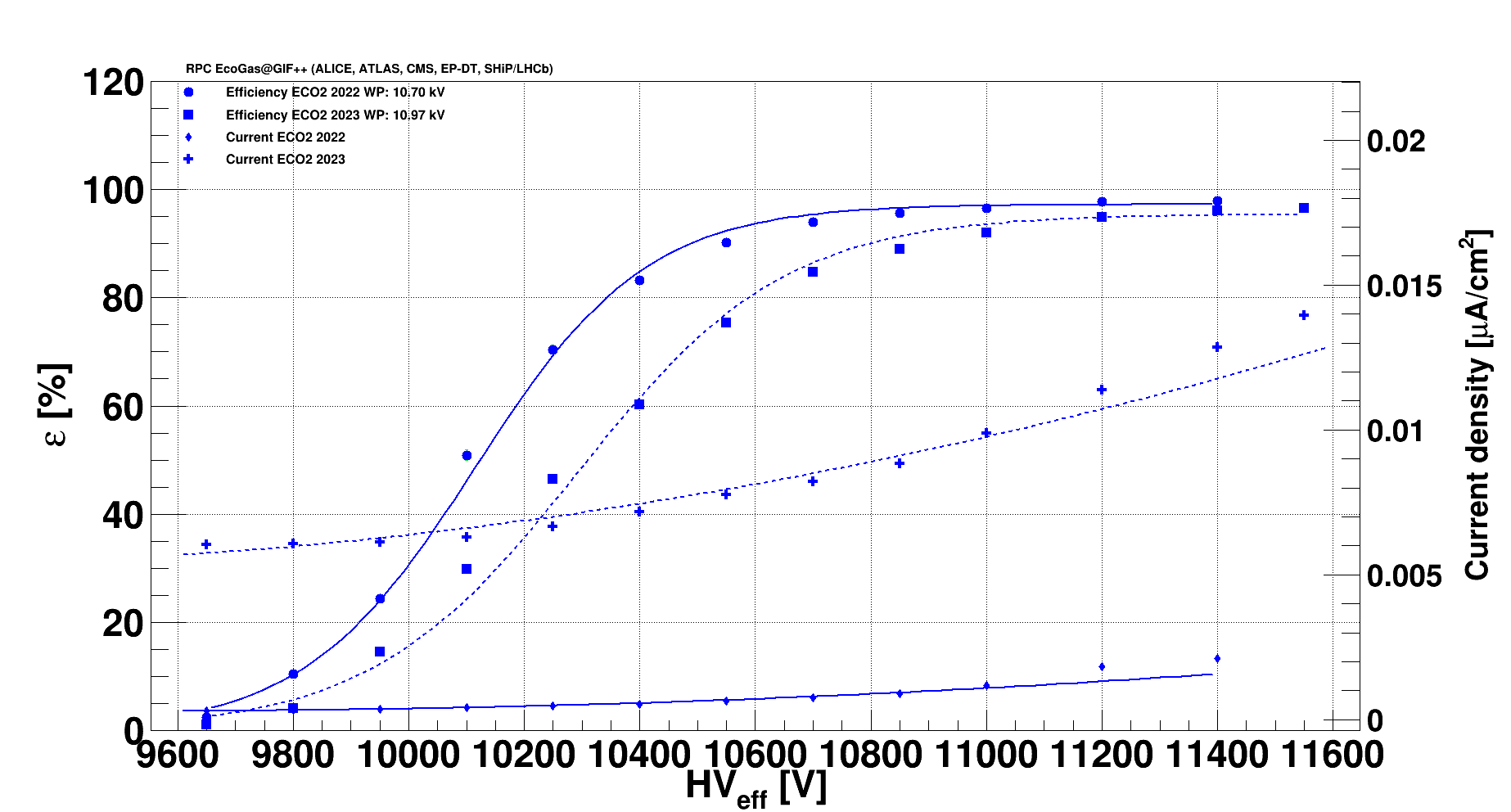}}
 \label{fig:Alice_ECO2}
  \hfill
 \subfigure{\includegraphics[width=0.329\linewidth]{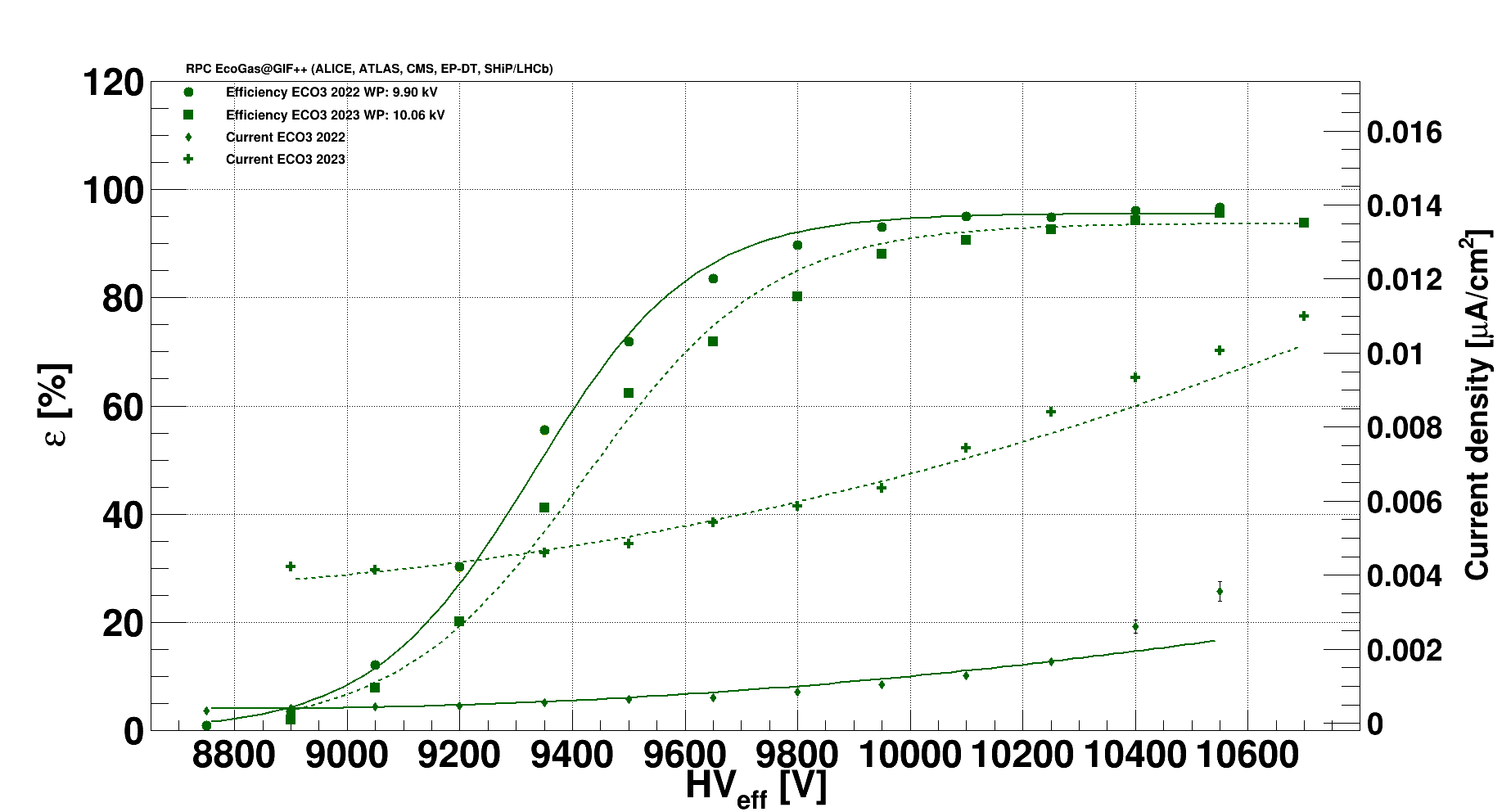}}\label{fig:Alice_ECO3}
  \caption{Left: Efficiency and current density versus $HV_{\text{eff}}$ comparing 2022 and 2023 data for the ALICE chamber operated with the STD mixture, obtained without irradiation. Center: Corresponding results for the same chamber filled with the ECO2 mixture. Right: Same, but with the same chamber filled with the ECO3 mixture.}
  \label{fig:side_by_side5}
\end{figure}

To spot potential aging, the RPC performance during and after irradiation was compared to baseline measurements~\cite{abbrescia2025long}. For source-off conditions, Figure~\ref{fig:side_by_side5} shows efficiency and current density as a function of $\mathrm{HV}_{\mathrm{eff}}$ for the ALICE RPC measured in 2022 and 2023. Plateau efficiencies for the STD, ECO2, and ECO3 mixtures remain stable within uncertainties, but the efficiency curves are shifted towards higher operating voltages in 2023, which is consistent with the increased current density.
% Figure 8

\begin{figure}[H]
\centering
\includegraphics[width=.5\textwidth]{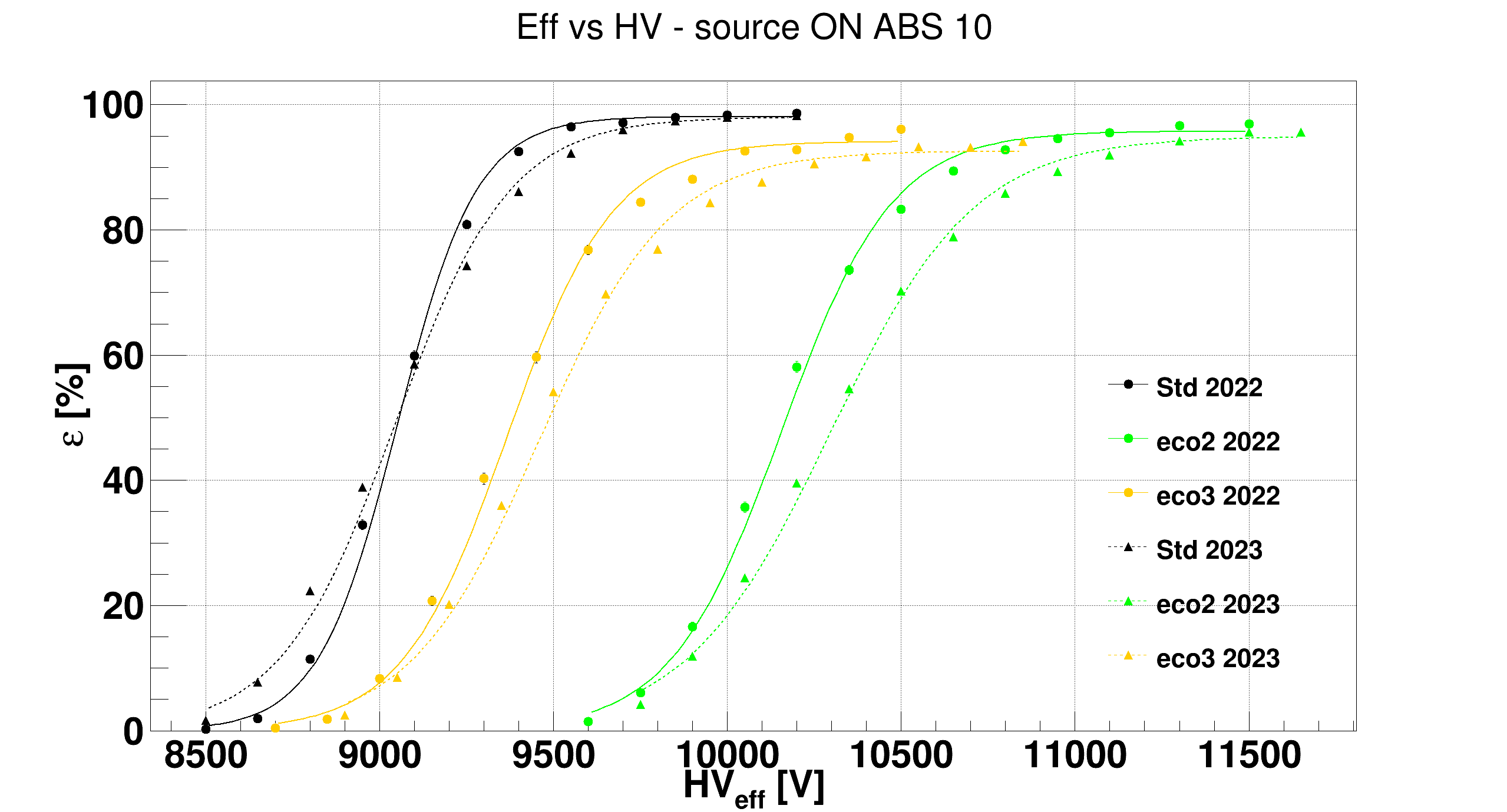}
\caption{Efficiency curves of ALICE RPC with STD, ECO2, and ECO3 gas mixtures in 2022 and 2023 under \ce{^{137}Cs} irradiation (ABS = 10).} 
% After one year of irradiation, a slight reduction in plateau efficiency and a shift to higher operating voltages are observed.
\label{fig:Alice_on_after}
\end{figure}

\begin{figure}[htbp]
\centering
\subfigure[]{\includegraphics[width=0.35\linewidth]{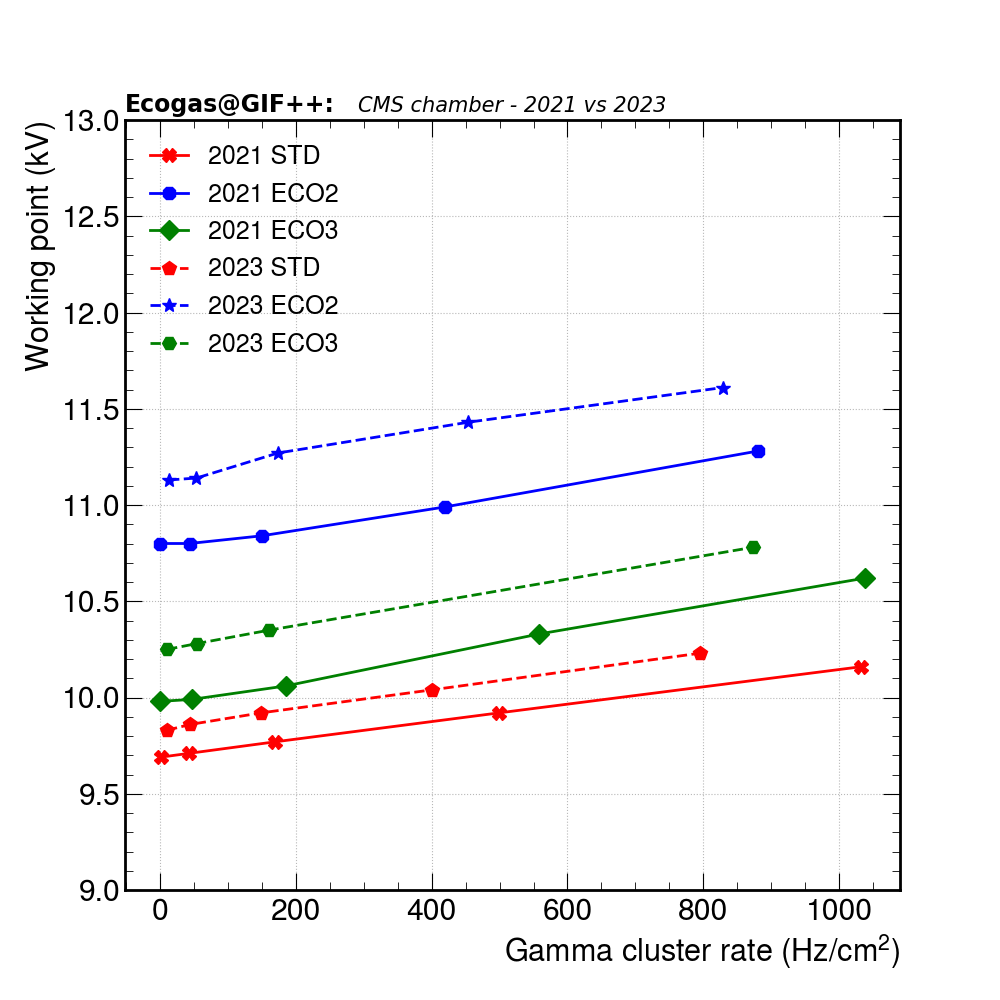}
\label{fig:RE11_WP_gamma}}
\hspace{0.05\linewidth}
\subfigure[]{\includegraphics[width=0.35\linewidth]{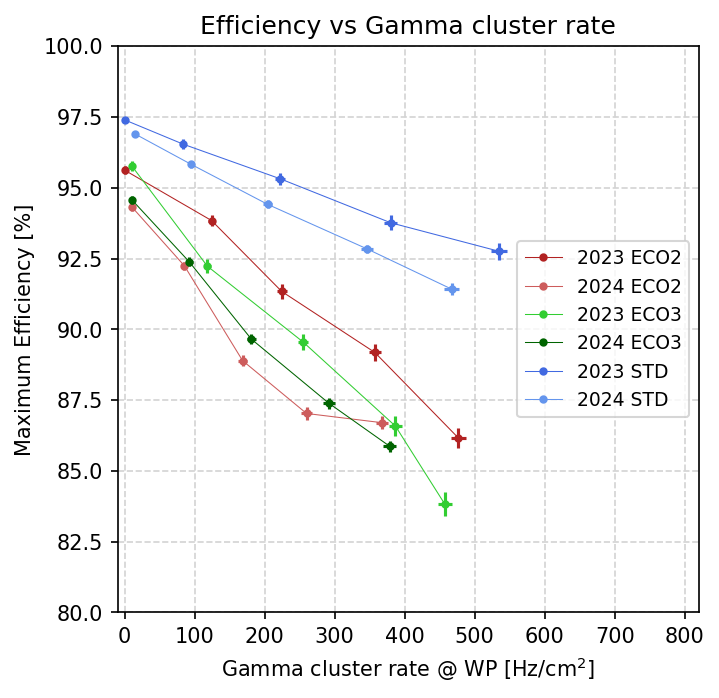}
\label{fig:EPDT_Max_eff}}
\caption{(a): Working point voltage as a function of gamma cluster rate for CMS RPC and different gas mixtures obtained with CMS RPC.(b): Maximum efficiency as a function of gamma cluster rate for EP-DT RPC measured in 2023 and 2024.}
\label{fig:side_by_side}
\end{figure}

Under irradiation with the \ce{^{137}Cs} source (ABS = 10), the efficiency curves as a function of $\mathrm{HV}_{\mathrm{eff}}$ measured in 2023 were compared to those obtained in 2022 for the ALICE RPC, as shown in Figure~\ref{fig:Alice_on_after}. For the STD, ECO2, and ECO3 mixtures, the 2023 data exhibit only a slight reduction in plateau efficiency after one year of irradiation. Nonetheless, for all gas mixtures, the efficiency curves are systematically shifted towards higher voltage values (after an accumulated charge of approximately 80 mC/cm$^2$) compared to the 2022 data.  This may reflect itself in higher values of current density at larger absorbed dose. This effect arises from the intrinsic resistive nature of the RPC electrodes and the consequent voltage drop across them that keeps the current flowing from their outer to their inner surfaces and results in the fact that the voltage across the gas gap ($HV_{\text{gas}}$) is effectively lower than the applied voltage, by an amount that can be computed using Ohm's law. For ECO2 and ECO3, this drop is larger than for STD, reflecting their higher current densities.
% Figure 9
Figure~\ref{fig:RE11_WP_gamma} shows the dependence of the CMS RPC working point on the gamma radiation background level, comparing baseline measurements from 2021 with data taken in 2023, after two years of irradiation. In both data sets, the working point is seen to increase with higher gamma background for all mixtures. However, the values from 2023 are consistently shifted towards higher voltages, likely due to the integrated charge accumulated over the two years of irradiation. The shift is more pronounced for the ECO2 and ECO3 mixtures compared to the STD one, possibly due to the higher current densities involved, as discussed in (Section~\ref{sec:perb}). Figure~\ref{fig:EPDT_Max_eff} shows the maximum efficiency of the EP-DT RPC as a function of the gamma cluster rate in 2023 and 2024, where the efficiency is seen to decrease with increasing irradiation, most notably for ECO2 in 2024. These effects are under investigation and require further study to confirm the underlying mechanisms. 
% Figure 10
% \FloatBarrier 

\section{Conclusion}
The performance and long-term aging of RPCs with eco-friendly gas mixtures are being studied at the CERN GIF++ facility by the RPC ECOgas@GIF++ Collaboration. Baseline measurements indicated that with the ECO2 and ECO3 gas mixtures, RPC efficiencies above 95\% are reachable, but at the cost of higher operating voltages and current densities compared to the standard gas mixture. Under irradiation, the plateau efficiencies decrease and the chamber working points shift to higher voltage values, with stronger effects for the eco-friendly mixtures. After a few years in the aging campaign, stable efficiencies are maintained, although the observed gradual increases in dark and ohmic currents suggest a change in electrode resistivity. The results demonstrate the feasibility of eco-friendly mixtures for RPC operation under HL-LHC–like conditions, but point to the need for further optimization studies to mitigate the higher currents and working voltage shifts.

\label{sec:con}

\appendix

\acknowledgments
This work is supported by Vrije Universiteit Brussel via OZR project \#OZR4149, and by the EU Horizon 2020 Research and Innovation program under grant agreement No.\ 101004761.

% Bibliography

% Recommended: using JHEP.bst file
\bibliographystyle{JHEP}
\bibliography{sources.bib}

\providecommand{\href}[2]{#2}\begingroup\raggedright\begin{thebibliography}{10}

\bibitem{santonico1981development}
R.~Santonico and R.~Cardarelli, \emph{Development of resistive plate counters}, {\emph{Nuclear Instruments and Methods in Physics Research} {\bfseries 187} (1981) 377}.

\bibitem{cms1997electromagnetic}
{CMS Collaboration}, \emph{The electromagnetic calorimeter project},  Technical Design Report CERN-LHCC-97-033; CMS-TDR-4, CERN (1997).

\bibitem{cortese1999alice}
{ALICE Collaboration}, \emph{Alice dimuon forward spectrometer},  Technical Design Report CERN-LHCC-99-022; ALICE-TDR-5, CERN (1999).

\bibitem{lacava1997atlas}
{ATLAS Collaboration}, \emph{Atlas muon spectrometer},  Technical Design Report CERN-LHCC-97-022; ATLAS-TDR-10, CERN (1997).

\bibitem{EU517}
{EU Council}, ``{Regulation 2014/517}.'' \url{https://eur-lex.europa.eu/eli/reg/2014/517/oj}, 2014.

\bibitem{no2024573}
{EU Council}, ``{Regulation 2024/573}.'' \url{https://eur-lex.europa.eu/eli/reg/2024/573/oj/eng}, 2024.

\bibitem{ecogaspaper2024}
L.~Quaglia et~al., \emph{Eco-friendly resistive plate chambers for detectors in future hep applications}, {\emph{Nucl.Instrum.Meth.A} {\bfseries 1058} (2024) 168757}.

\bibitem{schmidt2016high}
B.~Schmidt, \emph{The high-luminosity upgrade of the lhc: Physics and technology challenges for the accelerator and the experiments},  in \emph{Journal of Physics: Conference Series}, vol.~706, p.~022002, IOP Publishing, 2016.

\bibitem{abbrescia2013operation}
M.~Abbrescia et~al., \emph{Operation, performance and upgrade of the cms resistive plate chamber system at lhc}, {\emph{Nuclear Instruments and Methods in Physics Research Section A: Accelerators, Spectrometers, Detectors and Associated Equipment} {\bfseries 732} (2013) 195}.

\bibitem{abbrescia2024high}
M.~Abbrescia et~al., \emph{High-rate tests on resistive plate chambers operated with eco-friendly gas mixtures}, {\emph{The European Physical Journal C} {\bfseries 84} (2024) 300}.

\bibitem{abbrescia2025long}
M.~Abbrescia et~al., \emph{Long-term ageing studies on eco-friendly resistive plate chamber detectors}, {\emph{Particles} {\bfseries 8} (2025) 15}.

\end{thebibliography}\endgroup

\end{document}